\documentclass[twocolumn,showpacs,preprintnumbers,amsmath,amssymb]{revtex4}
\input epsf
\usepackage{graphicx}
\usepackage{dcolumn}
\usepackage{bm}

\begin{document}
\preprint{}

\title{Rayleigh Scattering and Atomic Dynamics in Dissipative Optical Lattices}

\author{F.-R. Carminati, L. Sanchez-Palencia, M. Schiavoni,
        F. Renzoni and G. Grynberg}
\address{Laboratoire Kastler-Brossel, D\'epartement de Physique de l'Ecole
Normale Sup\'erieure, 24, rue Lhomond, 75231, Paris Cedex 05,
France.}

\date{\today}

\begin{abstract}
We investigate Rayleigh scattering in dissipative optical lattices.
In particular, following recent proposals (S.~Guibal {\it et al}, Phys.
Rev. Lett. {\bf 78}, 4709 (1997); C.~Jurczak {\it et al}, Phys. Rev. Lett.
{\bf 77}, 1727 (1996)), we study whether the Rayleigh resonance originates
from the diffraction on a density grating, and is therefore a probe of
transport of atoms in optical lattices. It turns out that this is not the
case: the Rayleigh line is instead a measure of the cooling rate, while
spatial diffusion contributes to the scattering spectrum with a much broader
resonance.
\end{abstract}


\maketitle

Light scattering \cite{boyd}, i.e. the scattering of photons resulting from
the interaction with a material medium, is a technique widely used to
determine the properties of many different types of media.
From the position and the width of the scattering resonances it is
in fact possible to identify the dynamical modes of the system and derive
the rates of relaxation toward equilibrium. This is well exemplified by
the Landau-Placzek relation, valid for light scattering originating
from the density fluctuations of a medium at thermal equilibrium, which
connects the strength of the different components of the scattering spectrum
to the specific heats of the medium at constant volume and constant pressure.

Recently light scattering has been extensively used to study the properties
of cold atomic samples, and in particular it turned out to be an essential 
tool for the understanding of the basic properties of dissipative optical
lattices \cite{robi}. The same technique may also apply to far-off-resonance
non-dissipative optical lattices which are currently investigated by many
groups in connection with Bose-Einstein condensation experiments \cite{bec}.
However to derive the damping rates of the system from light scattering 
measurements is in general a highly non-trivial task. This is especially true 
for quasi-elastic (Rayleigh) scattering \cite{andreas,analog,aspect}, which 
gives access to the relaxation rates of non-propagating material observables.
In this work we investigate the mechanism behind the Rayleigh scattering in
dissipative optical lattices, and identify the relaxation process which 
determines the width of the Rayleigh resonance in the scattering spectrum.

The starting point of the present study is the previous claim that Rayleigh
resonances may originate from the excitation of the atomic density, and 
consequently the width of the Rayleigh line would provide a measure of the 
diffusion coefficients of the atoms in an optical lattice \cite{analog}. 
Following a similar approach, Jurczak {\it et al} \cite{aspect} derived 
values for the diffusion coefficients from polarization-selective intensity 
correlations. 

In our analysis we first assume, on the lines of these previous works, that 
the material observable excited in the pump-probe spectroscopy is the atomic 
density, and derive the expected relation between the width of the Rayleigh 
resonance and the spatial diffusion coefficients. Through experimental and 
theoretical work we show that this relation is actually {\it not} satisfied
by independent measurements/calculations of the width of the resonance and 
the diffusion coefficients. Instead, we show that the narrow Rayleigh resonance
originates from the atomic velocity damping, i.e. the width of the resonance
is a measure of the cooling rate, while spatial diffusion contributes to the
scattering spectrum with a much broader resonance.

Consider first the general relation between the width of the Rayleigh line
and the relaxation rate of the material observable excited in the optical 
process. In the basic setup of pump-probe spectroscopy, an atomic sample 
interacts with two laser fields: a strong pump beam, with frequency $\omega$, 
and a weak probe beam with frequency $\omega + \delta$. The superimposition 
of the pump and probe fields results in an interference pattern moving with 
phase velocity $v=\delta / |\Delta \vec{k}|$, with $\Delta \vec{k}$ the 
difference between pump and probe wavevectors. The atomic sample tends to
follow the interference pattern and a grating of an atomic observable
(typically density, magnetization or temperature) is created.
However due to the finite response time of the atomic medium the material
grating is phase-shifted with respect to the light interference pattern.
Therefore the pump beam can be diffracted on the material grating
in the direction of the probe, modifying the probe transmission.
It is then clear that it should be possible to derive information about the
atomic response time from the transmission spectrum. More precisely if we 
assume that only one atomic observable is excited in the optical process,  
and that the time evolution of this observable is characterized by a single
relaxation rate $\gamma$, the probe gain spectrum $g(\delta)$ has then a 
dispersive line shape
\begin{equation}
g\propto \frac{\delta}{\gamma^2+\delta^2}
\end{equation}
with peak-to-peak distance $2\gamma$, as derived in \cite{courtois96bis}.

Consider now the specific configuration with linearly-polarized pump and probe 
beams, the two polarization vectors being parallel. The resulting {\it 
intensity} interference pattern gives rise, via the dipole force, to a grating 
of the atomic density $n$ of the form
\begin{equation}
n=n_0 +n_1 ( \exp [ -i(\delta\cdot t-\Delta\vec{k}\cdot \vec{r})]
+c.c.) ~ .
\label{density}
\end{equation}
We assume now, following previous work \cite{analog}, that the Rayleigh 
resonance originates from the scattering on this {\it atomic density} grating.
It follows that the width of the Rayleigh line is related to the spatial 
diffusion coefficients. Indeed the relaxation mechanism of a grating of 
atomic density is spatial diffusion: atoms have to move to destroy the 
density grating. More quantitatively, if we assume that the dynamics of the
atoms in the optical lattice is well described by the Fick law
\begin{equation}
\frac{\partial n}{\partial t} = D_{x} \frac{\partial^2 n}{\partial x^2} + 
D_{y}\frac{\partial^2 n}{\partial y^2}+D_{z}\frac{\partial^2 n}{\partial z^2}~,
\label{fick}
\end{equation}
where $D_i$ ($i=x,y,z$) is the spatial diffusion coefficient in the $i$
direction, we find substituting the expression (\ref{density}) for $n$ in 
(\ref{fick}) that the relaxation rate $\gamma_D$ of the atomic density, 
defined by
\begin{equation}
\frac{\partial n_1}{\partial t} = -\gamma_D n_1~,
\label{relax}
\end{equation}
is given by
\begin{equation}
\gamma_D = D_x \Delta k_x^2 + D_y \Delta k_y^2 + D_z \Delta k_z^2~.
\label{main}
\end{equation}
Under the assumption that the Rayleigh resonance originates from the
scattering on the atomic density grating, the half-distance peak-to-peak of
the Rayleigh line $\gamma_R$ is simply equal to the relaxation rate $\gamma_D$,
and therefore measurement of $\gamma_R$ allows the determination of the
diffusion coefficients, as in  Refs. \cite{analog,aspect}. The validity of 
this approach will be tested by comparing results for the relaxation rate
$\gamma_D$ with measurements of the width of the Rayleigh resonance, as 
presented below.

\begin{figure}[ht]
\begin{center}
\mbox{\epsfxsize 2.5in \epsfbox{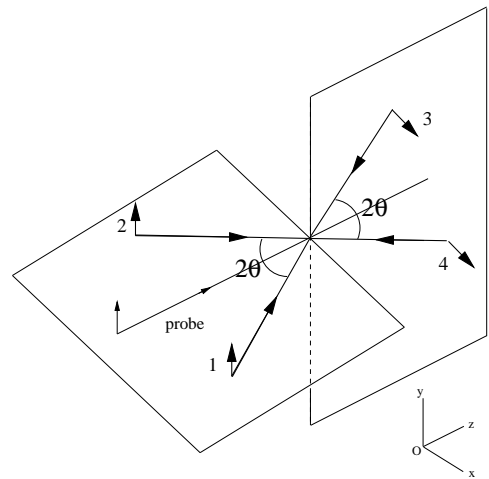}}
\end{center}
\vspace{-0.6cm}
\caption{Sketch of the experimental setup.}
\label{fig1}
\end{figure}

In our experiment rubidium atoms are cooled and trapped in a three-dimensional
(3D) lin$\perp$lin near resonant optical lattice \cite{robi}. The periodic 
structure is determined by the interference of four linearly polarized laser 
beams, arranged as in Fig. \ref{fig1}. The angle $2\theta$ between
copropagating 
lattice beams is equal to $60^0$. This four-beam configuration is the same, 
except for the value of the angle $\theta$, as the one considered in the works 
of Guibal {\it et al} \cite{analog} and Jurczak {\it et al} \cite{aspect}.

\begin{figure}[ht]
\begin{center}
\mbox{\epsfxsize 3.in \epsfbox{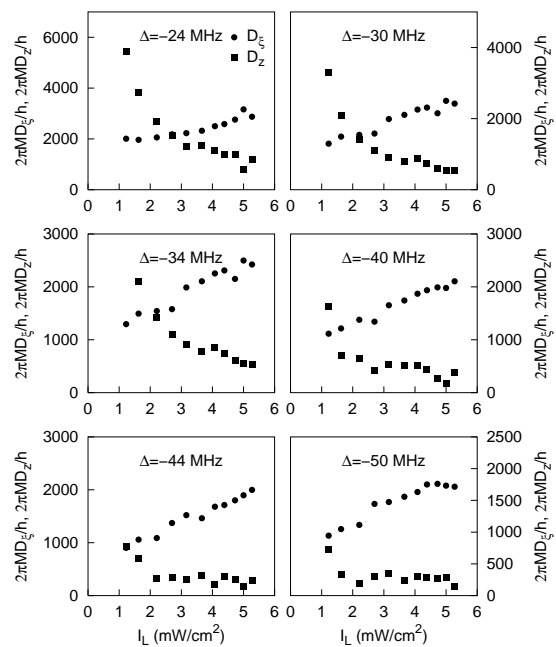}}
\end{center}
\vspace{-0.6cm}
\caption{Experimental results for the spatial diffusion coefficients in
the $\xi$ and $z$ directions as functions of the intensity per lattice
beam $I_L$ and for different values of the lattice detuning $\Delta$.}
\label{fig2}
\end{figure}

To determine {\it in a direct way} the spatial diffusion coefficients of
the atoms in the optical lattice, we observe the atomic cloud expansion
by using a Charge Coupled Device (CCD) camera \cite{hodapp,guidoni,epj1}.
Since the $x$ and $y$ directions are equivalent in our lattice (see Fig.
\ref{fig1}), we chose to take images in the $\xi Oz$ plane, where $\xi$ is
the axis in the $xOy$ plane forming an angle of $45^0$ with the $x$ and $y$
axis. Correspondingly, we determined the diffusion coefficients $D_{\xi}$
and $D_z$ in the $\xi$ and $z$ directions, with results as in Fig. \ref{fig2}.

These values for the diffusion coefficients are not consistent with the
value of about $10$ $\hbar/M$ determined for the same configuration by
Jurczak {\it et al} \cite{aspect} by polarization-selective intensity
correlations. As we will show in the following, this unconsistency derives
from the unreliability of the determination of the diffusion coefficients
by light scattering measurements, as this derivation of the diffusion
coefficients is based on the assumption that the narrow Rayleigh resonance
originates from the diffraction on an atomic density grating.

We turn now to the measurements of the width of the Rayleigh resonance.
The $y$ polarized probe beam is derived from the lattice beams, with the
relative detuning controlled with acousto-optical modulators. This probe
beam is sent along the $z$ axis through the cold atomic sample,
(Fig. \ref{fig1}) with its frequency scanned around the lattice beams'
frequency. The probe can interfere with the different lattice beams, which
play the role of the pump.

\begin{figure}[ht]
\begin{center}
\mbox{\epsfxsize 3.in \epsfbox{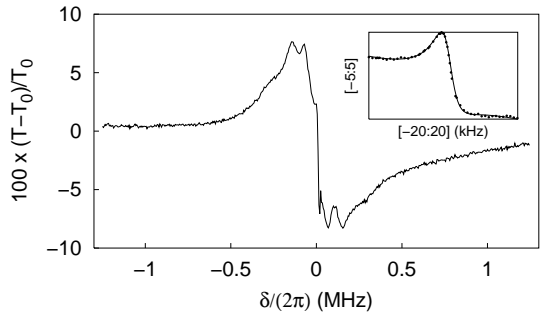}}
\end{center}
\vspace{-0.6cm}
\caption{Probe transmission as a function of the detuning between probe
and lattice beams, $T$ and $T_0$ being the intensity of the transmitted
probe beam with and without the atomic cloud. The inset shows a slow scan
of the region around zero detuning, together with the fit with the function
(\protect\ref{eq:fit}) (solid line).}
\label{fig3}
\end{figure}

A typical probe transmission spectrum is shown in Fig. \ref{fig3}. The lateral
resonances have been characterized in great detail in past investigations
\cite{brillo02}, and we focus here on the resonance at the center of the
spectrum (inset of Fig. \ref{fig3}). To determine whether this Rayleigh
resonance can be associated with the relaxation mechanism of spatial diffusion,
we made a systematic study of the width of the resonance as a function of the
interaction parameters (lattice-field intensity and detuning).

\begin{figure}[ht]
\begin{center}
\mbox{\epsfxsize 3.5in \epsfbox{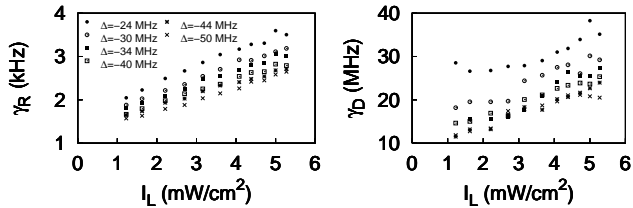}}
\end{center}
\vspace{-0.6cm}
\caption{Left: Experimental results for the half of the peak-to-peak distance
of the Rayleigh resonance. Right: Relaxation rate $\gamma_D$ of the atomic
density, as calculated from the experimental data for $D_{\xi}$ and $D_z$
using Eq.~(\protect\ref{gamma_th}). Both quantities are plotted as functions
of the intensity per lattice beam, for different values of the lattice detuning.}
\label{fig4}
\end{figure}

The peak-to-peak distance $2\gamma_R$ of the Rayleigh resonance has been
determined by fitting the central part of the probe transmission spectrum
with the function
\begin{equation}
f(\delta)=a_1+a_2\delta+\frac{a_3}{\delta^2+\gamma^2_R}+
\frac{a_4\delta}{\delta^2+\gamma^2_R}~.
\label{eq:fit}
\end{equation}
Here the linear term in detuning describes the wings of the sideband
resonances \cite{brillo02}. The Lorentzian resonance arises from the radiation
pressure, and has the same width of the dispersive line, as discussed in Ref.
\cite{analog}. Experimental results for $\gamma_R$ are reported in Fig.
\ref{fig4} as a function of the lattice beam intensity, for different values
of the lattice detuning.

We now describe the determination of $\gamma_D$. In the examined configuration
the probe beam can interfere simultaneously with all lattice beams. Therefore
the situation is slightly more complicated than the one analyzed previously
leading to Eq. \ref{main}, and to derive the link between the width of the
Rayleigh resonance and the diffusion coefficients we have to calculate the
interference pattern between the probe and the lattice beams. By using the
expression for the lattice-beams electric fields for a 3D lin$\perp$lin
optical lattice \cite{robi}, we easily find that the intensity-modulation
produced by the probe beam is:
\begin{equation}
\delta|\vec{E}|^2 \simeq E_p^{*} E_0 \cos(K x)
                    \exp\{ i [ (K_{+}-k)z+\delta\cdot t] \} + c.c.~,
\end{equation}
with $E_0$ ($E_p$) the amplitude of the lattice (probe) field,
$K=k\sin\theta$ and $K_{+}=k\cos\theta$.
Substituting in Fick's law, Eq. (\ref{fick}), the resulting modulation
for the atomic density we find that the relaxation rate $\gamma_D$,
defined via Eq. (\ref{relax}), is in the present case
\begin{equation}
\gamma_D = D_x (k \sin\theta)^2 +  D_z k^2 (1-\cos\theta)^2~.
\label{gamma_th}
\end{equation}
This equation is consistent with the relation derived in Ref.
\cite{analog} in the limit of small $\theta$. To determine whether the
rate $\gamma_D$ of relaxation of the atomic density is equal to the width
of the Rayleigh resonance, we calculate from the values $D_{\xi}, D_z$ of
Fig.  \ref{fig2} the relaxation rate $\gamma_D$ of the atomic density, using
Eq. (\ref{gamma_th}), with results as in Fig. \ref{fig4}.  For the same range
of interaction parameters the relaxation rate $\gamma_D$ is four orders of
magnitude larger than the half-distance peak-to-peak $\gamma_R$ of the
Rayleigh resonance.  We therefore conclude that the Rayleigh resonance does
not originate from the diffraction on an atomic density grating, and therefore
measurements of the width of the Rayleigh line do not allow the determination
of the spatial diffusion coefficients.

\begin{figure}[ht]
\begin{center}
\mbox{\epsfxsize 3.3in \epsfbox{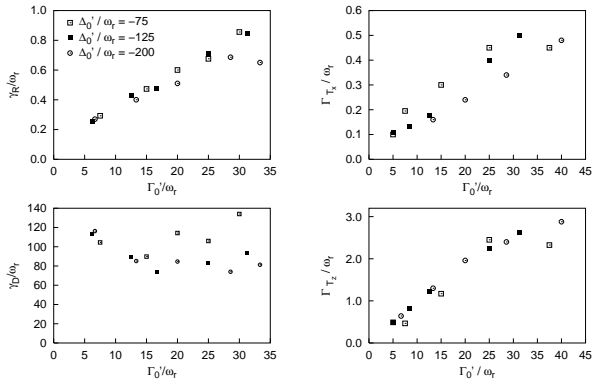}}
\end{center}
\vspace{-0.6cm}
\caption{Numerically calculated relaxation rate of the atomic density
$\gamma_D$, half-distance peak-to-peak $\gamma_R$ of the Rayleigh line
and relaxation rates $\Gamma_{T_x}$, $\Gamma_{T_z}$ of the atomic
temperature in the $x$ and $z$ directions. All these quantities are reported
as functions of the optical pumping rate $\Gamma_0'$, for different values
of the light shift per beam $\Delta_0'$. Here $\omega_r$ is the atomic recoil
frequency.}
\label{fig5}
\end{figure}

Our conclusions, based on the presented experimental findings, are supported
by numerical calculations. We consider a $J_g=1/2 \to J_e=3/2$ atomic
transition, as customary in numerical analysis of Sisyphus cooling. Taking
advantage of the symmetry between the $x$ and $y$ directions (see Fig.
\ref{fig1}), we restricted the atomic dynamics to the $xOz$ plane. Through
semiclassical Monte Carlo calculations \cite{petsas,sanchez}, we simulate the
dynamics of the atoms in the optical lattice. From the atomic trajectories we
determine then the probe
transmission spectra and the evolution of the atomic mean square displacements.
We calculate the width of the Rayleigh line and the spatial diffusion
coefficients.  From these diffusion coefficients we then derive through Eq.
(\ref{gamma_th}) the relaxation rate of the atomic density. The comparison
between the numerically calculated $\gamma_R$ and $\gamma_D$, as shown in
Fig.~\ref{fig5}, confirms that the width of the Rayleigh line does not
correspond to the rate of relaxation of the atomic density.

The final step of our analysis consists in identifying the damping process
which leads to the phase shift producing the Rayleigh scattering. Inspired 
by previous studies of stimulated Rayleigh scattering in corkscrew optical
molasses \cite{cork}, we numerically examined the damping process of the 
atomic velocity in the optical lattice and calculated the relaxation rates
$\Gamma_{T_x}$, $\Gamma_{T_z}$ of the atomic temperature in the $x$ and $z$
directions, with results as in Fig. \ref{fig5}. It appears that the damping 
rates of the atomic temperature not only are of the same order of magnitude
of the width of the Rayleigh line, but that they also display the same linear 
dependence on the optical pumping rate $\Gamma_0'$, at fixed light shift per
beam $\Delta_0'$, i.e. at fixed depth of the potential wells. More precisely,
neglecting $\Gamma_{T_x}$ as $\Gamma_{T_x}<<\Gamma_{T_z}$ we find from 
the data of Fig. \ref{fig5} that 
\begin{equation}
\gamma_R= 0.13(\pm 0.04)\omega_R+0.25 (\pm 0.02) \Gamma_{T_z}~.
\end{equation}
This shows that for an optical lattice the width of the Rayleigh line is a 
measure of the cooling rate, a behaviour already encountered in corkscrew 
optical molasses. It is then legitimate to investigate the eventual 
contribution of the light scattering on the density grating to the probe
transmission spectrum. By fitting the broad wings of the
numerically calculated spectra with a dispersive function, we found that the
corresponding width is approximately equal to the relaxation rate 
$\gamma_D$. This shows that the scattering on the density grating contributes
to the probe transmission spectrum with a resonance much broader than the
narrow line observed at the center of the spectra. In other
words, the information on the spatial diffusion coefficients is contained in 
the broad wings ($\omega\gtrsim $ 10 MHz) of the scattering spectrum, and not 
in the central narrow resonance.

In summary, in this work we investigated the connection between Rayleigh 
scattering and the atomic dynamics in dissipative optical lattices.
In particular, following recent proposals \cite{analog,aspect}, we studied
whether the Rayleigh resonance originates from the diffraction on a density
grating, and is therefore a probe of transport of atoms in optical lattices. 
It turns out that this is not the case: the Rayleigh line is instead a measure
of the cooling rate, while spatial diffusion contributes to the scattering 
spectrum with a much broader resonance.

We thank David Lucas for comments on the manuscript. This work was supported
by R\'egion Ile de France under contract E.1220. Laboratoire Kastler Brossel 
is an "unit\'e mixte de recherche de l'Ecole Normale Sup\'erieure et de 
l'Universit\'e Pierre et Marie Curie associ\'ee au Centre National de la 
Recherche Scientifique (CNRS)".

\end{document}